# Triality and Bagger-Lambert Theory

Hitoshi Nishino[1]  and  Subhash Rajpoot[2]

*Department of Physics & Astronomy*
*California State University*
*1250 Bellflower Boulevard*
*Long Beach, CA 90840*

## Abstract

We present two alternative field contents for Bagger-Lambert theory, based on the triality of $SO(8)$. The first content is $\left(\varphi_{Aa}, \chi_{\dot{A}a}; A_\mu{}^{ab}\right)$, where the bosonic field $\varphi$ is in the $\mathbf{8}_\mathrm{S}$ of $SO(8)$ instead of the $\mathbf{8}_\mathrm{V}$ as in the original Bagger-Lambert formulation. The second field content is $\left(\varphi_{\dot{A}a}, \chi^I{}_a; A_\mu{}^{ab}\right)$, where the bosonic field $\varphi$ and the fermionic field $\chi$ are respectively in the $\mathbf{8}_\mathrm{C}$ and $\mathbf{8}_\mathrm{V}$ of $SO(8)$. In both of these field contents, the bosonic potentials are positive definite, as desired. Moreover, these bosonic potentials can be unified by the triality of $SO(8)$. To this end, we see a special constant matrix as a product of two $SO(8)$ generators playing an important role, relating the $\mathbf{8}_\mathrm{V}$, $\mathbf{8}_\mathrm{S}$ and $\mathbf{8}_\mathrm{C}$ of $SO(8)$ for the triality. As an important application, we give the supersymmetry transformation rule for $N=6$ superconformal Chern-Simons theory with the supersymmetry parameter in the $\mathbf{6}$ of $SO(6)$, obtained by the truncation of our first field content.

PACS: 11.30.Pb, 12.60.Jv, 11.25.Hf, 11.25.-w, 11.27.Yb

Key Words: Multiple M2-Branes, $N=8$ Extended Supersymmetry, Triality of $SO(8)$,
    Conformal Symmetry, Chern-Simons Terms, and Three Dimensions.

[1] E-Mail: hnishino@csulb.edu
[2] E-Mail: rajpoot@csulb.edu

1. Introduction

It has been recently pointed out by Bagger and Lambert (BL) [1][2] that the totally antisymmetric triple brackets or 3-Lie algebras [3][4]

$$\left[X^I, X^J, X^K\right] \equiv \tfrac{1}{3!} \left[\left[X^I, X^J\right], X^K\right] \pm \text{(cyclic perms.)} \tag{1.1}$$

for the element $X^I$ of non-associative algebra play a crucial role in the context of coincident M2-brane which in turn is one of the important aspects of M-theory [5][6]. In [1][2], an explicit lagrangian in three-dimensions (3D) with global $N = 8$ supersymmetry has been given with $SO(4)_{\text{local}} \times SO(8)_{\text{global}}$ symmetry and a Chern-Simons (CS) term.

Afterwards, BL theory [1][2] has induced many different directions of investigations. For example, $OSp(8|4)$ superconformal symmetry in BL theory [1][2] has been confirmed [7] with potential generalizations to more general algebras. The algebraic structure [3] of BL theory [1][2] has also been studied from the viewpoint of embedding tensor [8][9], or that of $SU(2) \times SU(2)$ instead of $SO(4)$ [10], Lie 3-algebra [11] and its Kac-Moody extension [12]. Many relationships have been explored, such as the ones between M2-branes and D2-branes [13][14], relationships with M-5 branes [15], or with holographic dual [16], or with M-folds [17], with $N = 6$ superconformal CS theory [18], with the conformal limit [19] of Aharony-Bergman-Jafferis-Maldacena (ABJM) theory [20], and also with Janus field theory [21]. The BPS states in BL theory have also been extensively studied [22]. Mass deformations of the BL theory have been considered with the breaking $SO(8) \to SO(4) \times SO(4)$ [23], one-parameter deformation with non-compact metric [24], or the breaking $N = 8 \to N = 1$ by octonion-based mass parameters [25]. Other new investigations triggered by BL theory [1][2] are such as getting $N = 4$ membrane action [26] or ABJM theory [20] *via* orbifolds [27], or getting the couplings of M-2 branes to antisymmetric fluxes [28]. BL theory [1][2] has also been reformulated in terms of $N = 1$ superfield [29], studied on the plane-wave background [30], and on the light-cone [31].

There have been further generalizations to arbitrary non-compact Lie algebras [14][32] whose ghost problem has been overcome by spontaneous conformal symmetry breaking [33]. However, the uniqueness of the gauge group $SO(4)_{\text{local}}$ has been confirmed in [34] at least for compact gauge groups. In any case, due to the tight $N = 8$ system [1][2] strictly constraining the field content, together with the uniqueness of $SO(4)_{\text{local}}$ [34], it seems extremely difficult to generalize or change the basic field content of the original BL theory [1][2].

In this paper, we address the last question, *i.e.*, whether the basic field content of BL theory [1][2] can be changed, or whether there is any alternative field content. Here by 'the field content of the original BL formulation', we mean the case when the $SO(4)_{\text{local}}$ gauge group is specified with the bosonic field $X^I{}_a$ and its fermionic partner $\psi_{Aa}$ as in [2]. As explicit examples, we provide two alternative field contents to the original BL formulation [1]. Our first alternative field content is $\left(\varphi_{Aa}, \chi_{\dot{A}a}; A_\mu{}^{ab}\right)$, where the boson $\varphi_{Aa}$ is in the $\mathbf{8}_S$ (spinorial) instead of the $\mathbf{8}_V$ (vectorial) of $SO(8)$ [1][2], while the fermion $\chi$ is in the $\mathbf{8}_C$ (conjugate-spinorial) of $SO(8)$. The spinor charge $Q_{\alpha I}$ is in the $\mathbf{8}_V$ of $SO(8)$ instead of the $\mathbf{8}_S$ in the original BL formulation [2]. Our second field content is $\left(\varphi_{\dot{A}a}, \chi^I{}_a; A_\mu{}^{ab}\right)$, where the boson $\varphi$ and fermion $\chi$ are respectively in the $\mathbf{8}_C$ and $\mathbf{8}_V$ of $SO(8)$. Correspondingly, the spinor charge $Q_{\alpha A}$ is in the $\mathbf{8}_S$ of $SO(8)$. These replacements are possible thanks to the triality among $\mathbf{8}_V$, $\mathbf{8}_S$ and $\mathbf{8}_C$ of $SO(8)$. We also show that our first field content with the supercharge in the $\mathbf{8}_V$ of $SO(8)$ has a direct link with $N=6$ CS-matter theory [20][35], in which the supercharge is in the $\mathbf{6}$ of $SO(6)$.

## 2. First Field Content

Our first field content is $\left(\varphi_{Aa}, \chi_{\dot{A}a}; A_\mu{}^{ab}\right)$, where the indices $A, B, \cdots = 1, 2, \cdots, 8$ are for the $\mathbf{8}_S$ of $SO(8)$, $\dot{A}, \dot{B}, \cdots = \dot{1}, \dot{2}, \cdots, \dot{8}$ are for the $\mathbf{8}_C$ of $SO(8)$, while $I, J, \cdots = 1, 2, \cdots, 8$ are for the $\mathbf{8}_V$ of $SO(8)$. The indices $a, b, \cdots = 1, 2, 3, 4$ are for the vectorial $\mathbf{4}$ of $SO(4)$. The indices $\mu, \nu, \cdots = 0, 1, 2$ for the 3D space-time with the signature $(\eta_{\mu\nu}) = \text{diag.}(-, +, +)$.

Our total action $I_1 \equiv \int d^3x\, \mathcal{L}_1$ for the first field content has the lagrangian[3]

$$\mathcal{L}_1 = -\tfrac{1}{2}(D_\mu \varphi_{Aa})^2 + \tfrac{1}{2}(\overline{\chi}_{\dot{A}a} \gamma^\mu D_\mu \chi_{\dot{A}a}) + \tfrac{1}{64} c^{-1} \epsilon^{\mu\nu\rho} \epsilon^{abcd}(F_{\mu\nu}{}^{ab} A_\rho{}^{cd} - \tfrac{2}{3} A_\mu{}^{ab} A_\nu{}^{ce} A_\rho{}^{ed})$$
$$+ \tfrac{1}{4} c\, \epsilon^{abcd}(\overline{\chi}_a \Gamma^{IJ} \chi_b)(\varphi_c \Gamma^{IJ} \varphi_d) - \tfrac{4}{3} c^2 (\epsilon^{abcd} \varphi_{Bb} \varphi_{Cc} \varphi_{Dd})^2 \ . \tag{2.1}$$

Since the bosonic field $\varphi$ is in the $\mathbf{8}_S$ of $SO(8)$, we use the expressions, such as the last line, *e.g.*, $(\Gamma^{IJ})_{AB} \equiv (\Gamma^{[I})_{A\dot{C}} (\Gamma^{J]})_{\dot{C}B}$ for $(\Gamma^I)_{\dot{A}B} = -(\Gamma^I)_{B\dot{A}}$, and

$$(\varphi_c \Gamma^{IJ} \varphi_d) \equiv \varphi_{Ac} (\Gamma^{IJ})_{AB} \varphi_{Bd} \ . \tag{2.2}$$

The $SO(4)$-covariant derivative $D_\mu$ acts on the $\varphi$'s and $\chi$'s as

$$D_\mu \varphi_{Aa} \equiv \partial_\mu \varphi_{Aa} + A_{\mu a}{}^b \varphi_{Ab} \ , \quad D_\mu \chi_{\dot{A}a} \equiv \partial_\mu \chi_{\dot{A}a} + A_{\mu a}{}^b \chi_{\dot{A}b} \ . \tag{2.3}$$

---

[3] We do not distinguish the superscript/subscripts for the $SO(4)$ indices $a, b, \cdots$ or $SO(8)$ indices $A, B, \cdots$; $\dot{A}, \dot{B}, \cdots$ and $I, J, \cdots$, due to their positive definite metrics for contractions. We sometimes use both of them in order to clarify the contractions, such as in (2.3) through (2.5).

In the last term in (2.1), the 'square' implies all the free indices $a$, $B$, $C$ and $D$ in one pair of the parentheses are contracted. This gives the manifestly positive-definite bosonic potential

$$V_1 \equiv +\tfrac{4}{3}c^2(\epsilon^{abcd}\varphi_{Bb}\varphi_{Cc}\varphi_{Dd})^2 \geq 0 \ . \tag{2.4}$$

This potential has an alternative expression given in (2.14). Compared with [2], our CS term is exactly the same as that in [2], and so is the positive definiteness of the bosonic potential [2], while the $\chi^2\varphi^2$ term has the same magnitude as that in [2].

Our physical field content $(\varphi_{Aa}, \chi_{\dot{A}a})$ is in a sense similar to $N=16$ $\sigma$-model with the coset $E_{8(+8)}/SO(16)$ [36][37]. Because the latter has the physical field content $(\varphi_A, \chi_{\dot{A}})$ with the index $A = 1, 2, \cdots, 128$ (or $\dot{A} = \dot{1}, \dot{2}, \cdots, \overline{128}$) in the **128** (or $\overline{\mathbf{128}}$) of $SO(16)$. In our notation, we do not need the imaginary unit '$i$' in front of the fermionic kinetic term, except that needed due to the signature $(+,-,-)$ in [37]. Due to the Clifford algebra structures repeated at every eight space-time dimensions [38], the $SO(8)$ spinorial structures of our system must be parallel to the case of $SO(16)$ in [37]. From this viewpoint, we adopt the notation with no imaginary unit in front of the $\chi$-kinetic term. Accordingly, we need *no* imaginary unit in front of the $\varphi$-kinetic term, either. The consistency of our notation will be seen as the emergence of the positive-definite potential (2.14a).

Our total action $I$ is invariant under the $SO(4)_{\text{local}}$ symmetry

$$\delta_G \varphi_{Aa} = -\alpha_a{}^b \varphi_{Ab} \ , \quad \delta_G \chi_{\dot{A}a} = -\alpha_a{}^b \chi_{\dot{A}b} \ ,$$
$$\delta_G A_\mu{}^{ab} = +D_\mu \alpha^{ab} \equiv +\partial_\mu \alpha^{ab} + A_\mu{}^{ac}\alpha_c{}^b + A_\mu{}^{bc}\alpha^a{}_c \ , \tag{2.5}$$

$SO(8)_{\text{global}}$ symmetry

$$\delta_H \varphi_{Aa} = -\tfrac{1}{4}\beta^{IJ}(\Gamma^{IJ})_{AB}\varphi_{Ba} \ , \quad \delta_H \chi_{\dot{A}a} = -\tfrac{1}{4}\beta^{IJ}(\Gamma^{IJ})_{\dot{A}\dot{B}}\chi_{\dot{B}a} \ , \quad \delta_H A_\mu{}^{ab} = 0 \ , \tag{2.6}$$

and global $N=8$ supersymmetry

$$\delta_Q \varphi_{Aa} = +(\Gamma^I)_{A\dot{B}}(\bar{\epsilon}^I \chi_{\dot{B}a}) \ ,$$
$$\delta_Q \chi_{\dot{A}a} = -(\Gamma^I)_{B\dot{A}}(\gamma^\mu \epsilon^I) D_\mu \varphi_{Ba} - \tfrac{2}{3}c\,\epsilon_a{}^{bcd}\,\epsilon^I(\Gamma^J \varphi_b)_{\dot{A}}(\varphi_c \Gamma^{IJ}\varphi_d) \ ,$$
$$\delta_Q A_\mu{}^{ab} = +4c\,\epsilon^{abcd}(\Gamma^I \varphi_c)_{\dot{B}}(\bar{\epsilon}^I \gamma_\mu \chi_{\dot{B}d}) \ . \tag{2.7}$$

Since $\varphi$ is in the $\mathbf{8}_{\text{S}}$ of $SO(8)$, we frequently use the expressions, *e.g.*, $(\Gamma^I \varphi_b)_{\dot{A}} \equiv (\Gamma^I)_{\dot{A}B}\varphi_{Bb}$. The structure of supersymmetry transformation (2.7) is parallel to that in the original formulation [1][2], such as the $D\varphi$ or $\varphi^3$-term in $\delta_Q\chi$, and $\chi\varphi$-term in $\delta_Q A_\mu$. However, the great difference is that now the supersymmetry parameter $\epsilon^I$ is in the $\mathbf{8}_V$ of $SO(8)$.

The closure of two supersymmetries works just as in the original formulation [2]. In fact, at the linear order, we have

$$[\delta_Q(\epsilon_1), \delta_Q(\epsilon_2)] = \delta_P(\xi_3) + \delta_G(\alpha_3) \ , \tag{2.8}$$

where $\delta_P$ is the translation with the parameter $\xi_3^\mu \equiv +2(\overline{\epsilon}_1^I \gamma^\mu \epsilon_2^I)$, while $\delta_G$ is the $SO(4)_{\text{local}}$ transformation with the parameter $\alpha_3^{ab} \equiv -\xi^\mu A_\mu{}^{ab}$. Compared with the original formulation [2], due to the supersymmetry parameter $\epsilon^I$ in the $\mathbf{8}_V$ of $SO(8)$, the explicit index $I$ is needed in $\xi_3^\mu$.

The positive definite potential $V_1$ and the $\varphi^3$-term in $\delta_Q\chi$ can be re-expressed in terms of the generalized 'superpotential' $W_{ABCD}$ as

$$W_{ABCD} \equiv + \tfrac{1}{24}\epsilon^{abcd} \varphi_{Aa} \varphi_{Bb} \varphi_{Cc} \varphi_{Dd} \ , \tag{2.9a}$$

$$V_1 = + \tfrac{768}{25} c^2 \left( \tfrac{\partial W_{ABCD}}{\partial \varphi_{Aa}} \right)^2 \geq 0 \ , \tag{2.9b}$$

$$\delta_Q \chi_{\dot{A}a} \Big|_{\varphi^3} = - \tfrac{16}{5} c \left(\Gamma^I\right)_{B\dot{A}} (\Gamma^{IJ})_{CD} \, \epsilon^J \left( \tfrac{\partial W_{ABCD}}{\partial \varphi_{Aa}} \right) \ . \tag{2.9c}$$

On the RHS of (2.9b), the index $A$ is contracted within the parentheses, while the indices $a, B, C, D$ are contracted, when the pair of parentheses is squared.

The positive definiteness of our potential is a non-trivial conclusion. Because it is the reflection of the total consistency of our system, such as the usage of our notation, in which both the fermionic and bosonic inner products do not have any imaginary unit '$i$' in front. This convention has been already used in $N = 16$ supergravity [37].

The confirmation of supersymmetry $\delta_Q I_1 = 0$ is more involved than the original formulation [2]. However, the basic cancellation in each sectors is parallel to [2]. In fact, the confirmation works as follows. At the *quadratic* order, the computation is routine. At the *cubic* order, we have only the $\chi F \varphi$-terms, which are parallel to [2].

At the *quartic* order, we have two sectors of terms: (i) $(D\chi)\varphi^3$ and (ii) $\chi^3 \varphi$. For the sector (i), we need the identity

$$A_{BC} \equiv +\tfrac{1}{16}(\Gamma^{IJ})_{BC}(\Gamma^{IJ})_{DE}\, A_{DE} \ , \tag{2.10}$$

for any antisymmetric tensor $A_{BC} = -A_{CB}$. It turns out that all the terms have only two structures

$$\epsilon^{abcd}(\Gamma^{IJK})_{A\dot{B}}(\bar{\epsilon}^I\gamma^\mu\chi_{\dot{B}b})(\varphi_c\Gamma^{JK}\varphi_d)D_\mu\varphi_{Aa} \ ,$$
$$\epsilon^{abcd}(\Gamma^I)_{A\dot{B}}(\bar{\epsilon}^J\gamma^\mu\chi_{\dot{B}b})(\varphi_c\Gamma^{IJ}\varphi_d)D_\mu\varphi_{Aa} \ . \tag{2.11}$$

The conditions of vanishing of these two kinds of terms determine the coefficients of the $\chi^2\varphi^2$-term in the lagrangian and of the $\varphi^3$-terms in $\delta_Q\chi$.

In the sector (ii) $\chi^3\varphi$, we have three different structures of terms:[4]

$$(A) \equiv +\epsilon^{abcd}(\bar{\epsilon}^K\Gamma^K\Gamma^{IJ}\chi_b)_A(\bar{\chi}_c\Gamma^{IJ}\chi_d)\varphi_{Aa} \ , \tag{2.12a}$$

$$(B) \equiv +\epsilon^{abcd}(\bar{\epsilon}^I\gamma_\mu\Gamma^I\Gamma^{[4]}\chi_b)_A(\bar{\chi}_c\gamma^\mu\Gamma^{[4]}\chi_d)\varphi_{Aa} \ , \tag{2.12b}$$

$$(C) \equiv +\epsilon^{abcd}(\bar{\epsilon}^I\gamma_\mu\Gamma^I\chi_b)_A(\bar{\chi}_c\gamma^\mu\chi_d)\varphi_{Aa} \ . \tag{2.12c}$$

However, as the Fierzing of each of $(A)$, $(B)$ and $(C)$ reveals, there are two relationships among them:

$$(A) = -8(B) \ , \quad (C) = -240(B) \ . \tag{2.13}$$

Thus, all the terms no more than the $(B)$-terms, and their cancellation uniquely fixes the coefficient of the $\chi^2\varphi^2$-term in the lagrangian.

At the *quintic* order, there is no term arising as in [2]. However, at the final *sextic* order, there is one sector of the type $\chi\varphi^5$. The analysis of this sector needs special care. First, we note that the $\varphi^6$-term in $\mathcal{L}_1$ can be re-expressed as an alternative form

$$\mathcal{L}_{1,\varphi^6} = -V_1 \equiv -\tfrac{4}{3}c^2(\epsilon^{abcd}\varphi_{Bb}\varphi_{Cc}\varphi_{Dd})^2 \tag{2.14a}$$

$$\equiv -\tfrac{7}{8}c^2(\varphi_a\varphi_a)^3 + \tfrac{1}{128}c^2(\varphi_a\varphi_a)(\varphi_b\Gamma^{IJKL}\varphi)^2$$
$$+ \tfrac{1}{768}c^2(\varphi_a\Gamma^{IJKL}\varphi_a)(\varphi_b\Gamma^{KLMN}\varphi_b)(\varphi_c\Gamma^{MNIJ}\varphi_c) \ . \tag{2.14b}$$

Second, it turns out that all the terms in the *sextic* order fall in one of the following four structures (1P), (1Q), (3P) and (5P) defined by

$$(1P) \equiv (\bar{\epsilon}^L\chi_{\dot{C}b})(\Gamma^L\varphi_b)_{\dot{C}}(\varphi_c\Gamma^{IJ}\varphi_d)^2 = +\tfrac{7}{6}(\xi) - \tfrac{1}{96}(\kappa) \ , \tag{2.15a}$$

---
[4] We use the symbol $\Gamma^{[n]}$ for totally antisymmetric $\Gamma$-indices. For example, $\Gamma^{[4]}$ stands for $\Gamma^{KLMN}$.

$$(1Q) \equiv (\overline{\epsilon}^L \chi_{\dot{C}b})(\Gamma^J \varphi_b)_{\dot{C}} (\varphi_c \Gamma^{LK} \varphi_d)(\varphi_c \Gamma^{KJ} \varphi_d) = -\tfrac{7}{48}(\xi) + \tfrac{1}{768}(\kappa) \ , \qquad (2.15\text{b})$$

$$(3P) \equiv (\overline{\epsilon}^L \chi_{\dot{C}b})(\Gamma^{JMN} \varphi_b)_{\dot{C}} (\varphi_c \Gamma^{JL} \varphi_d)(\varphi_c \Gamma^{MN} \varphi_d) = +\tfrac{1}{128}(\eta) + \tfrac{1}{768}(\zeta) \ , \qquad (2.15\text{c})$$

$$(5P) \equiv (\overline{\epsilon}^L \chi_{\dot{C}b})(\Gamma^L \Gamma^{IJMN} \varphi_b)_{\dot{C}} (\varphi_c \Gamma^{IJ} \varphi_d)(\varphi_c \Gamma^{MN} \varphi_d) = -\tfrac{1}{16}(\eta) - \tfrac{1}{96}(\zeta) \ , \qquad (2.15\text{d})$$

where the terms $(\xi)$, $(\eta)$, $(\zeta)$ and $(\kappa)$ are defined by

$$(\xi) \equiv \delta_Q \left[ (\varphi_a \varphi_a)^3 \right] \ , \qquad (\eta) \equiv \left[ \delta_Q \left\{ (\varphi_a \Gamma^{[4]} \varphi)^2 \right\} \right] (\varphi_b \varphi_b) \ ,$$

$$(\zeta) \equiv \delta_Q \left[ (\varphi_a \Gamma^{IJKL} \varphi_a)(\varphi_b \Gamma^{KLMN} \varphi_b)(\varphi_c \Gamma^{MNIJ} \varphi_c) \right] \ , \qquad (\kappa) \equiv [\delta_Q (\varphi_a \varphi_a)] (\varphi_b \Gamma^{[4]} \varphi_a)^2 \ . \quad (2.16)$$

The lemmas in (2.15) can be easily obtained by Fierzing. The second expressions in (2.15a) and (2.15d) are straightforward, but those in (2.15b) and (2.15c) are non-trivial to get. The expressions in terms of $(\xi)$, $(\eta)$, $(\zeta)$ and $(\kappa)$ are convenient to integrate to compare $\delta_Q \mathcal{L}_{1,\varphi^6}$. In particular, the coefficient of the terms $(\eta)$ and $(\kappa)$ out of $\delta_Q \mathcal{L}_{1,\chi^2 \varphi^2}$ should be the same for them to be cancelled by $\delta_Q \mathcal{L}_{1,\varphi^6}$.

## 3. Second Field Content

Our second field content is $\left( \varphi_{\dot{A}a}, \chi^I{}_a; A_\mu{}^{ab} \right)$. Other than the representational difference of fields, the index convention is exactly the same as in section 2, e.g., $\varphi$ in the $\mathbf{8}_\text{C}$ and $\chi$ in the $\mathbf{8}_\text{V}$ of $SO(8)$. The lagrangian for our total action $I_2 \equiv \int d^3 x \, \mathcal{L}_2$ is

$$\begin{aligned}
\mathcal{L}_2 = &-\tfrac{1}{2}(D_\mu \varphi_{\dot{A}a})^2 + \tfrac{1}{2}(\overline{\chi}^I{}_a \gamma^\mu D_\mu \chi^I{}_a) \\
&+ \tfrac{1}{64} c^{-1} \epsilon^{\mu\nu\rho} \epsilon^{abcd} (F_{\mu\nu}{}^{ab} A_\rho{}^{cd} - \tfrac{2}{3} A_\mu{}^{ab} A_\nu{}^{ce} A_\rho{}^{ed}) \\
&+ c \, \epsilon^{abcd} (\overline{\chi}^I{}_a \chi^J{}_b)(\varphi_c \Gamma^{IJ} \varphi_d) - \tfrac{4}{3} c^2 (\epsilon^{abcd} \varphi_{\dot{B}b} \varphi_{\dot{C}c} \varphi_{\dot{D}d})^2 \ .
\end{aligned} \qquad (3.1)$$

Since the $\varphi$'s is in the $\mathbf{8}_\text{C}$ of $SO(8)$, we have the expressions, such as $(\varphi_c \Gamma^{IJ} \varphi_d) \equiv \varphi_{\dot{A}c} (\Gamma^{IJ})_{\dot{A}\dot{B}} \varphi_{\dot{B}d}$. Our action $I_2$ is invariant under $SO(8)_\text{global}$, $SO(4)_\text{local}$ and global $N = 8$ supersymmetry

$$\delta_Q \varphi_{\dot{A}a} = + (\Gamma^I)_{B\dot{A}} (\overline{\epsilon}_B \chi^I{}_a) \ , \qquad (3.2\text{a})$$

$$\delta_Q \chi^I{}_a = - (\Gamma^I)_{A\dot{B}} (\gamma^\mu \epsilon_A) D_\mu \varphi_{\dot{B}a} + \tfrac{2}{3} c \, \epsilon^{abcd} \epsilon_A (\Gamma^J \varphi_b)_A (\varphi_c \Gamma^{IJ} \varphi_d) \ , \qquad (3.2\text{b})$$

$$\delta_Q A_\mu{}^{ab} = + 4 c \, \epsilon^{abcd} (\Gamma^I \varphi_c)_A (\overline{\epsilon}_A \gamma_\mu \chi^I{}_d) \ . \qquad (3.2\text{c})$$

Here again, we are using the notations, such as $(\Gamma^I \varphi_b)_A \equiv (\Gamma^I)_{A\dot{B}} \varphi_{\dot{B}b}$. The supersymmetry parameter $\epsilon_A$ is now in the $\mathbf{8}_\text{S}$ of $SO(8)$.

The closure of supersymmetries works just as in our first field content and the original formulation [2] as well. At the linear order, we have

$$[\delta_Q(\epsilon_1), \delta_Q(\epsilon_2)] = \delta_P(\xi_3) + \delta_G(\alpha_3) \ , \tag{3.3}$$

with $\xi_3^\mu \equiv +2(\overline{\epsilon}_1 \gamma^\mu \epsilon_2)$ for the translation $\delta_P$, and $\alpha_3^{ab} \equiv -\xi^\mu A_\mu{}^{ab}$ for the $SO(4)_{\text{local}}$ transformation $\delta_G$. The supersymmetry parameter $\epsilon_A$ now is in the $\mathbf{8}_{\text{S}}$ of $SO(8)$, so that the index $A$ is suppressed in $\xi_3^\mu$.

Also in our second field content, its bosonic potential $V_2 \equiv -\mathcal{L}_{2,\varphi^6}$ is positive definite:

$$V_2 \equiv +\tfrac{4}{3} c^2 \left( \epsilon^{abcd} \varphi_{\dot{B}b} \varphi_{\dot{C}c} \varphi_{\dot{D}d} \right)^2 \geq 0 \ . \tag{3.4}$$

The coefficient $4c^2/3$ is the same as in the original formulation [1]. The bosonic potential $V_2$ and the $\varphi^3$-term in $\delta_Q \chi$ can be re-expressed in terms of the generalized superpotential $W_{\dot{A}\dot{B}\dot{C}\dot{D}}$ as

$$W_{\dot{A}\dot{B}\dot{C}\dot{D}} \equiv +\tfrac{1}{24} \epsilon^{abcd} \varphi_{\dot{A}a} \varphi_{\dot{B}b} \varphi_{\dot{C}c} \varphi_{\dot{D}d} \ , \tag{3.5a}$$

$$V_2 = +\tfrac{768}{25} c^2 \left( \frac{\partial W_{\dot{A}\dot{B}\dot{C}\dot{D}}}{\partial \varphi_{\dot{A}a}} \right)^2 \geq 0 \ , \tag{3.5b}$$

$$\delta_Q \chi^I{}_a \Big|_{\varphi^3} = +\tfrac{16}{5} c \, (\Gamma^J)_{A\dot{B}} (\Gamma^{IJ})_{\dot{C}\dot{D}} \, \epsilon_A \left( \frac{\partial W_{\dot{A}\dot{B}\dot{C}\dot{D}}}{\partial \varphi_{\dot{A}a}} \right) \ . \tag{3.5c}$$

These structures are parallel to the first field content case in (2.9).

The invariance confirmation $\delta_Q I_2 = 0$ is very parallel to $\delta_Q I_1 = 0$. Even the lemmas in (2.15) are parallel. For example, (2.15a) is simply replaced by

$$\widetilde{(1P)} \equiv (\overline{\epsilon}_A \chi^L{}_b)(\Gamma^L \varphi_b)_A (\varphi_c \Gamma^{IJ} \varphi_d)^2 = +[(\delta_Q \varphi_b) \varphi_b](\varphi_c \Gamma^{IJ} \varphi_d)^2 \ , \tag{3.6}$$

whose final form is eventually the same as in (2.15a), despite the different index assignments on the $\epsilon$'s, $\chi$'s and $\varphi$'s. Due to this parallel-ness, the confirmation of $\delta_Q I_2 = 0$ is greatly simplified.

Once we start performing the confirmation $\delta_Q I_2 = 0$, we see that the computation for the second field content is much easier than the first one. This is caused by the fact that the fermion $\chi^I{}_a$ is no longer in the $\mathbf{8}_{\text{C}}$, but in the $\mathbf{8}_{\text{V}}$ of $SO(8)$, so that necessary Fierzings are simpler.

### 4. Unification by Triality of $SO(8)$

We mention how the triality of $SO(8)$ works for the three formulations, *i.e.*, the original formulation in [2], and our first and second field contents.

First of all, we define the following constant $N$-matrices as products of two $SO(8)$ generators:

$$N^{IJKL}{}_{ABCD} \equiv \tfrac{1}{16}(\Gamma^{[IJ|})_{[AB|}(\Gamma^{|KL]})_{|CD]} \ , \qquad N^{IJKL}{}_{\dot{A}\dot{B}\dot{C}\dot{D}} \equiv \tfrac{1}{16}(\Gamma^{[IJ|})_{[\dot{A}\dot{B}|}(\Gamma^{|KL]})_{|\dot{C}\dot{D}]} \ . \quad (4.1)$$

These constant matrices play a central role in demonstrating the triality of $SO(8)$. For example, this constant matrix satisfies the (anti)self-duality conditions

$$N^{IJKL}{}_{ABCD} = -\tfrac{1}{24}\epsilon^{IJKL}{}_{MNPQ} N^{MNPQ}{}_{ABCD} \ , \tag{4.2a}$$

$$N^{IJKL}{}_{\dot{A}\dot{B}\dot{C}\dot{D}} = +\tfrac{1}{24}\epsilon^{IJKL}{}_{MNPQ} N^{MNPQ}{}_{\dot{A}\dot{B}\dot{C}\dot{D}} \ , \tag{4.2b}$$

$$N^{IJKL}{}_{ABCD} = -\tfrac{1}{24}\epsilon_{ABCD}{}^{EFGH} N^{IJKL}{}_{EFGH} \ , \tag{4.2c}$$

$$N^{IJKL}{}_{\dot{A}\dot{B}\dot{C}\dot{D}} = -\tfrac{1}{24}\epsilon_{\dot{A}\dot{B}\dot{C}\dot{D}}{}^{\dot{E}\dot{F}\dot{G}\dot{H}} N^{IJKL}{}_{\dot{E}\dot{F}\dot{G}\dot{H}} \ , \tag{4.2d}$$

with clear symmetries among these relationships, reflecting the triality between the $\mathbf{8}_{\rm V}$, $\mathbf{8}_{\rm S}$ and $\mathbf{8}_{\rm C}$ of $SO(8)$. Other important relationships are[5]

$$N^{IJKL}{}_{ABCD} N^{MNPQ}{}_{ABCD} = -\tfrac{1}{48}\epsilon^{IJKLMNPQ} + \tfrac{1}{2}\delta_I{}^{[M}\delta_J{}^N\delta_K{}^P\delta_L{}^{Q]} \ , \tag{4.3a}$$

$$N^{IJKL}{}_{ABCD} N^{IJKL}{}_{EFGH} = -\tfrac{1}{48}\epsilon_{ABCDEFGH} + \tfrac{1}{2}\delta_A{}^{[E}\delta_B{}^F\delta_C{}^G\delta_D{}^{H]} \ , \tag{4.3b}$$

$$N^{IJKL}{}_{\dot{A}\dot{B}\dot{C}\dot{D}} N^{MNPQ}{}_{\dot{A}\dot{B}\dot{C}\dot{D}} = +\tfrac{1}{48}\epsilon^{IJKLMNPQ} + \tfrac{1}{2}\delta_I{}^{[M}\delta_J{}^N\delta_K{}^P\delta_L{}^{Q]} \ , \tag{4.3c}$$

$$N^{IJKL}{}_{\dot{A}\dot{B}\dot{C}\dot{D}} N^{IJKL}{}_{\dot{E}\dot{F}\dot{G}\dot{H}} = -\tfrac{1}{48}\epsilon_{\dot{A}\dot{B}\dot{C}\dot{D}\dot{E}\dot{F}\dot{G}\dot{H}} + \tfrac{1}{2}\delta_{\dot{A}}{}^{[\dot{E}}\delta_{\dot{B}}{}^{\dot{F}}\delta_{\dot{C}}{}^{\dot{G}}\delta_{\dot{D}}{}^{\dot{H}]} \ . \tag{4.3d}$$

The proof of (4.2c) and (4.2d) can be simplified, if we use (4.3b) and (4.3d) by expressing the epsilon tensor in terms of the products of $\Gamma$-matrices. To our knowledge, these relationships associated with the triality of $SO(8)$ have never been explicitly given in the past.

If we compare the three potentials, *i.e.*, that in the original [2] and ours $V_1$ and $V_2$, they reveal the symmetric expressions for these three potentials:

$$V_0 = +\tfrac{4}{3}c^2(\epsilon^{abcd}\varphi^I{}_a\varphi^J{}_b\varphi^K{}_c)^2 = +\tfrac{32}{15}c^2\left(\epsilon^{abcd}N^{IJKL}{}_{ABCD}\varphi^I{}_a\varphi^J{}_b\varphi^K{}_c\right)^2 \ , \tag{4.4a}$$

$$V_1 = +\tfrac{4}{3}c^2(\epsilon^{abcd}\varphi_{Aa}\varphi_{Bb}\varphi_{Cc})^2 = +\tfrac{32}{15}c^2(\epsilon^{abcd}N^{IJKL}{}_{ABCD}\varphi_{Aa}\varphi_{Bb}\varphi_{Cc})^2 \ , \tag{4.4b}$$

$$V_2 = +\tfrac{4}{3}c^2\left(\epsilon^{abcd}\varphi_{\dot{A}a}\varphi_{\dot{B}b}\varphi_{\dot{C}c}\right)^2 = +\tfrac{32}{15}c^2\left(\epsilon^{abcd}N^{IJKL}{}_{\dot{A}\dot{B}\dot{C}\dot{D}}\varphi_{\dot{A}a}\varphi_{\dot{B}b}\varphi_{\dot{C}c}\right)^2 \ . \tag{4.4c}$$

Here $V_0$ is the bosonic potential in [2], and $\varphi^I{}_a$ is their $X^I{}_a$ in our notation. In (4.4), all the un-contracted indices within the pair of parentheses should be contracted when the pair of parentheses is squared. For example in (4.4b), the indices $d, I, J, K, L$ and $D$ are contracted, when the pair of parentheses is squared. Due to the second terms in (4.3), these give the desired symmetric expressions in the last sides of (4.4). In other words, we have a unified expression for (4.4) as

$$V = +\tfrac{32}{15}c^2 \left(\epsilon^{abcd}\mathcal{N}^{XYZU}{}_{X'Y'Z'U'}\,\varphi_{X'a}\varphi_{Y'b}\varphi_{Z'c}\right)^2 \quad, \tag{4.5}$$

where $\mathcal{N}$ stands for one of the three $N$'s in (4.4), depending on the representations of $\varphi_a$. For example, $\mathcal{N}^{XYZU}{}_{X'Y'Z'U'}$ implies $N^{IJKL}{}_{ABCD}$ for $\varphi_a$ in the $\mathbf{8}_\text{S}$ of $SO(8)$.

---

[5] Here we do not use the combination of the superscripts and subscripts for the contracted indices, because it is better to keep the order of $\mathbf{8}_{\rm V}$ superscripts and $\mathbf{8}_{\rm S}$ or $\mathbf{8}_{\rm C}$ subscripts for the matrix $N$. Also, for the products of Kronecker's deltas, we use the mixed indices for an obvious reason.

## 5. Relationships with $N=6$ Superconformal Chern-Simons Theory

As an important application of our first field content, we obtain the transformation rule for $N = 6$ superconformal Chern-Simons theory [20][35].[6]

The importance of this relationship stems from the fact that the supersymmetry parameter in our first field content is in the *vectorial* $\mathbf{8}_\text{V}$ of $SO(8)$, while the parameter for $N = 6$ theory is also in the *vectorial* $\mathbf{6}$ of $SO(6)$. By truncating the supersymmetry parameter in our first field content from the range of 8 into 6, we can reach the $N = 6$ theory [20][35]. In this process, we still keep the original $32 + 32$ degrees of freedom for physical fields. The difference from the recent works on $N = 6$ supersymmetry [20][35], however, is that the latters have $SU(\mathcal{N}) \times SU(\mathcal{N})$ or $U(\mathcal{N}) \times U(\mathcal{N})$ symmetry, while ours has only $SO(4)$.

The basic reduction rules are

$$\widehat{\Gamma}^{\hat{I}} = \begin{cases} \widehat{\Gamma}^i = \Gamma^i \otimes \sigma_1 & (i = 1, 2, \cdots, 6) \\ \widehat{\Gamma}^7 = \Gamma_7 \otimes \sigma_1 & , \\ \widehat{\Gamma}^8 = I_8 \otimes \sigma_2 & . \end{cases} \tag{5.1}$$

Here $\widehat{\Gamma}^{\hat{I}}$ are $16 \times 16$ antisymmetric matrices, including both chiralities for $SO(8)$, while *hats* are for $SO(8)$-related quantities and indices. The $\Gamma^i$'s are $8 \times 8$ antisymmetric $\gamma$-matrices for $SO(6)$ satisfying the usual Clifford algebra $\{\Gamma^i, \Gamma^j\} = +2\delta^{ij}$. As the number of components of $\Gamma^i$ shows, both chiralities, *i.e.*, $(\Gamma^i)_{\alpha\beta}$ and $(\Gamma^i)^{\alpha\beta}$ $(\alpha, \beta, \cdots = 1, 2, 3, 4)$ are represented by the $\Gamma^i$'s in (5.1). The $\Gamma_7$ is defined by $\Gamma_7 \equiv +i\Gamma^1\Gamma^2\cdots\Gamma^6$, controlling the chirality for $SO(6)$. Due to the peculiar structure of $SO(6) \approx SU(4)$, the subscript $\alpha$ and the superscript $\alpha$ respectively correspond to the positive and negative chiralities under $\Gamma_7$, and they are complex conjugations to each other. Accordingly, the chirality for $SO(8)$ corresponds to the eigen-states of the $\sigma_3$-matrix: $\widehat{\Gamma}^9 \equiv \widehat{\Gamma}^1\widehat{\Gamma}^2\widehat{\Gamma}^3\widehat{\Gamma}^4\widehat{\Gamma}^5\widehat{\Gamma}^6\widehat{\Gamma}^7\widehat{\Gamma}^8 = \sigma_3$. We also truncate $\epsilon^8 = \epsilon^9 = 0$, while maintaining our first field content with the original 32+32 degrees of freedom. Note that the symmetries of the both sides in (5.1) are consistent, because $\Gamma^i$ and $\Gamma_7$ are all antisymmetric.

---

[6] The special feature of $N = 6$ was pointed out also in locally superconformal theory [39].

Following this basic truncation rule, we can get the $N = 6$ transformation rule consistent with [20][35]

$$\delta_Q \varphi_{\alpha a} = + (\Gamma^i)_{\alpha\beta}(\bar{\epsilon}^i \chi^{*\beta}{}_a) \equiv +(\bar{\epsilon}^i \Gamma^i \chi^*{}_a)_\alpha \ , \tag{5.2a}$$

$$\delta_Q \varphi^{*\alpha}{}_a = + (\Gamma^i)^{\alpha\beta}(\bar{\epsilon}^i \chi_{\beta a}) \equiv +(\bar{\epsilon}^i \Gamma^i \chi_a)^\alpha \ , \tag{5.2b}$$

$$\delta_Q \chi_{\alpha a} = + (\gamma^\mu \Gamma^i \epsilon^i)_{\alpha\beta} D_\mu \varphi^{*\beta}{}_a + \tfrac{4}{3} c\,\epsilon^{abcd} \epsilon^i \varphi_{\alpha b}(\varphi_c^* \Gamma^i \varphi_d^*) - \tfrac{4}{3} c\,\epsilon^{abcd} \epsilon^i (\Gamma^j \varphi_b^*)_\alpha (\varphi_c^* \Gamma^{ij} \varphi_d) \ , \tag{5.2c}$$

$$\delta_Q \chi^{*\alpha}{}_a = - (\gamma^\mu \Gamma^i \epsilon^i)^{\alpha\beta} D_\mu \varphi_{\beta a} + \tfrac{4}{3} c\,\epsilon^{abcd} \epsilon^i \varphi^{*\alpha}{}_b(\varphi_c \Gamma^i \varphi_d) - \tfrac{4}{3} c\,\epsilon^{abcd} \epsilon^i (\Gamma^j \varphi_b)^\alpha (\varphi_c^* \Gamma^{ij} \varphi_d) \ , \tag{5.2d}$$

$$\delta_Q A_\mu{}^{ab} = + 4c\,\epsilon^{abcd}(\bar{\epsilon}^i \gamma_\mu \Gamma^i \chi_c^*)_\alpha \varphi^{*\alpha}{}_d + 4c\,\epsilon^{abcd}(\Gamma^i)^{\alpha\beta}(\bar{\epsilon}^i \gamma_\mu \Gamma^i \chi_c)^\alpha \varphi_{\alpha d} \ . \tag{5.2e}$$

We are using the notations, such as $(\varphi^*_c \Gamma^i \varphi^*_d) \equiv \varphi^{*\alpha}{}_c (\Gamma^i)_{\alpha\beta} \varphi^{*\beta}{}_d$, *etc*, to save space. The on-shell closure of gauge algebra is confirmed as

$$[\delta_Q(\epsilon_1), \delta_Q(\epsilon_2)] = +\delta_P(\xi_3^\mu) + \delta_G(\Lambda_3^{ab}) \ ,$$

$$\xi_3^\mu \equiv +2(\bar{\epsilon}_2 \gamma^\mu \epsilon_1) \ , \quad \Lambda_3^{ab} \equiv -\xi^\mu A_\mu{}^{ab} - 8c\,\epsilon^{abcd}(\bar{\epsilon}_1^i \epsilon_2^k)(\varphi_c^* \Gamma^{ik} \varphi_d) \ , \tag{5.3}$$

with the respective parameters $\xi^\mu$ and $\Lambda^{ab}$ for the translation and $SO(6)_{\text{local}}$ symmetry.

Up to the groups $SU(\mathcal{N}) \times SU(\mathcal{N})$ [20] and $U(\mathcal{N}) \times U(\mathcal{N})$ [35], which are replaced by $SO(4)$, our result is consistent with the $N = 6$ results [20][35]. For example, all of our transformations in (5.2) can be rewritten, such that our supersymmetry parameter $\epsilon^i$ appears only in the special combination $(\epsilon^i \Gamma^i)_{\alpha\beta}$ which can be identified with the supercharge $Q_{IJ}$ in eq. (2.8) in the second reference in [35].

## 6. Concluding Remarks

In this Letter, we have clarified the crucial role played by the triality of $SO(8)$ in BL theory [1][2]. Compared with the original formulation [1][2], our first field content $(\varphi_{Aa}, \chi_{\dot{A}a}; A_\mu{}^{ab})$ has the supersymmetry parameter $\epsilon^I$ in the $\mathbf{8}_V$ of $SO(8)$. Both the fermionic and bosonic fields are in the (conjugate) spinorial representations that is similar to the $N = 16$ maximal supersymmetric system in 3D [37]. As we have shown, this field content has a direct link with $N=6$ supersymmetry [20][35], where the supersymmetry parameter is also in the *vectorial* **6** of $SO(6)$. The second field content $\left(\varphi_{\dot{A}a}, \chi^I{}_a; A_\mu{}^{ab}\right)$ is complimentary to the first one, because the scalar field is now in the $\mathbf{8}_{\mathrm{C}}$ of $SO(8)$ that was not the case in the original [2] and our first field content.

Our scalar potentials in both formulations are positive definite, reflecting the total consistency of our system, such as the notation with the absence of the imaginary unit '$i$' in front of both fermionic and bosonic spinorial inner product. This convention has been already used in $N=16$ supergravity [37]. Reflecting the triality of $SO(8)$, the bosonic potentials $V_0$, $V_1$ and $V_2$ share exactly the same positive constant $4c^2/3$.

As has been mentioned in the Introduction, BL theory [1][2] can be obtained as the conformal limit of gauged supergravity [9]. From this viewpoint, our first content is the conformal limit of $N=8$ gauged supergravity with the physical fields $\left(\varphi_{Aa}, \chi_{\dot{A}a}\right)$ for the coset $SO(8,4)/SO(8)\times SO(4)$. Also, our second field content $\left(\varphi_{\dot{A}a}, \chi^I{}_a\right)$ can be obtained as the conformal limit of $N=8$ supergravity with the same coset, due to the triality of $SO(8)$.

We have also unified two potentials (2.4) and (3.4) by the triality of $SO(8)$ *via* the constant matrices $N^{IJKL}{}_{ABCD}$ and $N^{IJKL}{}_{\dot{A}\dot{B}\dot{C}\dot{D}}$. The three bosonic potentials $V_0$, $V_1$ and $V_2$ in the three formulations for different representations can be uniformly expressed in terms of the $\mathcal{N}$-matrix as in (4.5). As far as we know, these relationships have not been given explicitly in the context of $SO(8)$ triality in the past.

We have so far the three distinct formulations: the original BL theory with $\left(\varphi^I{}_a, \chi_{\dot{A}a}; \epsilon_A\right)$ [2], our first model with $\left(\varphi_{Aa}, \chi_{\dot{A}a}; \epsilon^I\right)$ and the second one with $\left(\varphi_{\dot{A}a}, \chi^I{}_a; \epsilon_A\right)$, where the $\epsilon$'s are supersymmetry parameters. Strictly speaking, there are three other formulations with $\left(\varphi_{\dot{A}a}, \chi_{Aa}; \epsilon^I\right)$, $\left(\varphi_{Aa}, \chi^I{}_a; \epsilon_{\dot{A}}\right)$ and $\left(\varphi^I{}_a, \chi_{Aa}; \epsilon_{\dot{A}}\right)$. However, the latter and the former are related through 'chirality-flip' conjugations with no essential differences.

Even though our field contents are natural consequences of $SO(8)$ triality, we emphasize that the new formulations of BL theory [1][2] presented here have not been entertained before. There are also many important applications, such as the truncation into $N=6$ supersymmetry [20][35].

This work is supported in part by NSF Grant # 0652996. We are indebted to the referee of this paper for the suggestion of giving an explicit connection between our first field content and $N=6$ theory [20][35].